\documentclass[final,3p,times]{elsarticle}
\usepackage{graphicx}
\usepackage{amssymb}
\journal{Physics Letters B}

\begin{document}

\begin{frontmatter}



\title{Supernova neutrino signals by liquid Argon detector and neutrino
magnetic moment}


\author[Yoshida]{Takashi Yoshida}
\author[Takamura]{Akira Takamura}
\author[Kimura]{Keiichi Kimura}
\author[Kawagoe]{Shio Kawagoe}
\author[Yoshida,Kajino]{Toshitaka Kajino}
\author[Kimura]{Hidekazu Yokomakura}
\address[Yoshida]{Department of Astronomy, Graduate School of Science, University of Tokyo, Tokyo 113-0033, Japan}
\address[Takamura]{Department of Mathematics, Toyota National College of Technology, Aichi 471-8525, Japan}
\address[Kimura]{Department of Physics, Graduate School of Science, Nagoya University, Aichi 464-8602, Japan}
\address[Kawagoe]{Knowledge Dissemination Unit, Institute of Industrial Science, University of Tokyo, Tokyo 153-8505, Japan}
\address[Kajino]{National Astronomical Observatory of Japan, Mitaka, Tokyo 181-8588, Japan}

\begin{abstract}
We study $\nu_e$ and $\bar{\nu}_e$ signals from a supernova with strong
magnetic field detected by a 100 kton liquid Ar detector.
The change of neutrino flavors by resonant spin-flavor conversions, matter 
effects, and neutrino self-interactions are taken into account.
Different neutrino signals, characterized by the total event numbers of
$\nu_e$ and $\bar{\nu}_e$ and neutronization burst event, 
are expected with 
different neutrino oscillation parameters and neutrino magnetic moment.
Observations of supernova neutrino signals by a 100 kton liquid Ar detector 
would constrain oscillation parameters as well as neutrino magnetic moment 
in either normal and inverted mass hierarchies.

\end{abstract}

\begin{keyword}
neutrinos \sep neutrino oscillations \sep supernovae \sep magnetic moment

\end{keyword}

\end{frontmatter}

% \linenumbers

\section{Introduction}
\label{}

Identification of very large neutrino magnetic moment 
($\mu_\nu \gg 10^{-18} \mu_B$, where $\mu_B$ is Bohr magnetons) is one of
important traces of particle physics beyond the standard model.
Neutrino experiments have constrained the upper limit of the
neutrino magnetic moment as $\mu_{\nu_e} < 5.8 \times 10^{-11} \mu_B$ 
\cite{be07}.
If neutrinos are Majorana particles and they have magnetic moment, 
they have only transition magnetic moment.
The magnetic moment enables a spin precession between a left-handed
neutrino $\nu_L$ and a right-handed antineutrino $\bar{\nu}_R$ with
different flavors in a strong magnetic field (e.g., \cite{lm88}).
This is the resonant spin-flavor (RSF) conversion.

Supernova (SN) neutrino is a good indicator observing the effects of 
neutrino magnetic moment.
The magnetic field of Fe core of a presupernova is expected to attain
$10^{10}$ gauss \cite{Heger05}.
Proto-neutron stars should have stronger magnetic field since observations of
pulsars have magnetic field of $\sim 10^{12}$ gauss.
If neutrinos are Majorana particles and have large transition magnetic
moment ($\mu_{\nu_{\alpha\beta}} \sim 10^{-12} \mu_B$), the RSF conversion 
is expected to occur in innermost region of SN ejecta 
(e.g., \cite{Ando03a,Ando03b,Ando03c,Akhmedov03}).
The final neutrino spectra also have influences of 
Mikheyev-Smirnov-Wolfenstein (MSW) effects (e.g., \cite{ds00}) and 
neutrino self-interactions (e.g., \cite{qf95}).
The influence of the RSF conversion, which will be observed in 
water-{\v C}erenkov detector, on SN $\bar{\nu}_e$ signal has been predicted
\cite{Ando03a,Ando03b,Ando03c,Yoshida09}.
The observations of $\bar{\nu}_e$ signals are expected to reveal 
the effects of the RSF conversions in inverted mass hierarchy.

A liquid Ar detector distinguishes $\nu_e$ and $\bar{\nu}_e$ events 
by charged-current (CC) reactions, events by neutral-current (NC) reactions 
and electron scattering of all flavors of neutrinos and antineutrinos.
Especially, many events of $\nu_e$ will be observed owing
to large cross section of $^{40}$Ar($\nu_e,e^-)$ reaction.
Previous studies have estimated the SN neutrino events by liquid Ar detector 
taking into account the MSW effects \cite{Gil-Botella03,Gil-Botella04} 
and neutrino self-interactions \cite{Choubey10}.
ICARUS experiment (e.g., \cite{Amoruso04}) is proposed as the usage of a
3 kton liquid Ar detector.
The experiment with the 600 ton detector T600 started at the underground 
laboratory in Gran Sasso \cite{Arneodo06}.
In Long-Baseline Neutrino Experiment Project it is proposed to construct 
three 17-kton liquid Ar detectors at DUSEL \cite{lbne10}.
Development of a very large liquid Ar detector with 100 kton is expected.

We investigate possible SN neutrino signals observed by a 100 kton liquid
Ar detector taking into account the RSF conversions,
MSW effects, and neutrino self-interactions.
We evaluate the SN neutrino events for $\nu_e$ and $\bar{\nu}_e$ by the CC
reactions and all flavors of neutrinos and antineutrinos by the NC reactions.
In this Letter, we report for the first time the possibility that the
effects of the RSF conversions and large neutrino magnetic moment are
extracted from SN signals in both normal and inverted mass
hierarchies.

\begin{figure}
\begin{center}
\includegraphics[width=7.5cm]{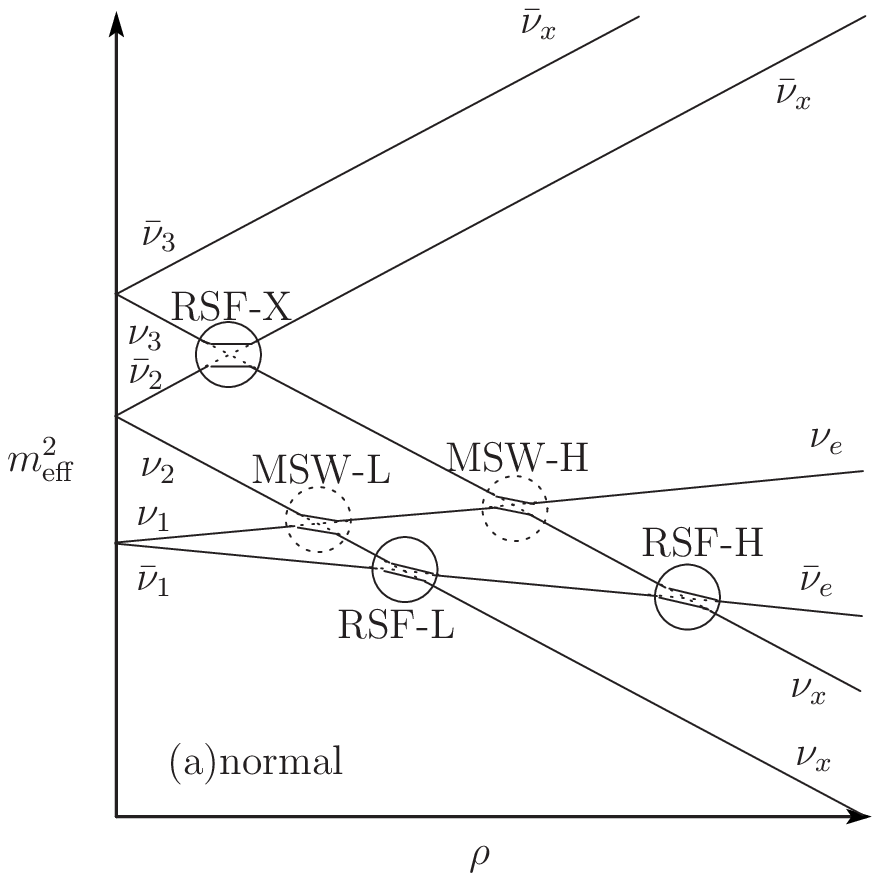}
\includegraphics[width=7.5cm]{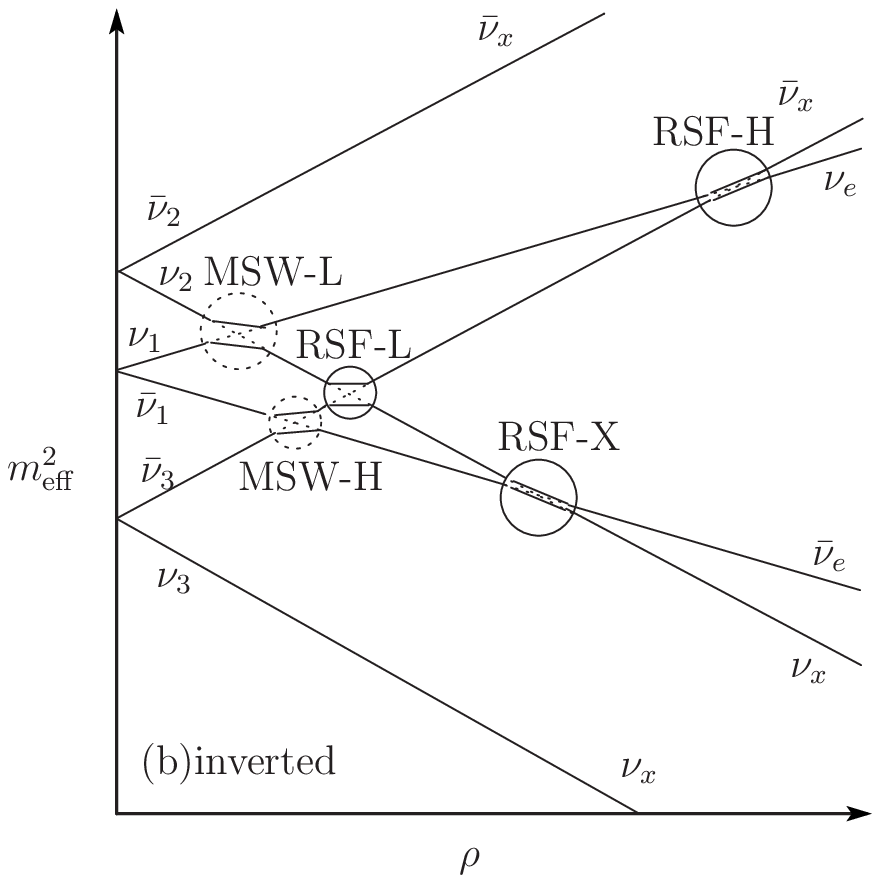}
\vspace*{-5mm}
\end{center}
\caption{
Schematic diagram of effective masses of neutrinos and antineutrinos
in (a) normal mass hierarchy and (b) inverted mass hierarchy.
Solid circles indicate three RSF resonances (RSF-H, RSF-X, and RSF-L) and 
dashed circles indicate two MSW resonances (MSW-H and MSW-L).
The changes of effective mass in accordance with a solid curve and dashed
curve indicate adiabatic and non-adiabatic conversions, respectively.
}
\end{figure}

\section{Model}

\subsection{SN neutrino model}

We used a SN model adopted in \cite{Yoshida09, Kawagoe10}.
The SN progenitor is evolved from a 15 $M_\odot$ star \cite{Woosley95}.
We used the SN density profiles at 0, 0.5, 1, 2, 3, 4, 5, 6, 7, 8, 9, and 10 s
after the core bounce \cite{Kawagoe10} to see the shock effect of the
RSF conversions and the MSW effects.
When the RSF conversions are taken into account, the SN magnetic field 
transverse to the neutrino propagation at the radius $r$
was assumed to be $B_{\perp}(r) = 10^{11} (10^8 {\rm cm}/r)^3$ gauss  
\cite{Ando03a,Yoshida09}.
The electron fraction in the innermost hot-bubble and wind region where
the mass coordinate is smaller than 1.43 $M_\odot$ was assumed to be 
$Y_e = 0.49$.
The electron fraction of the outer region was assumed to be the same as
that of the progenitor model.
We assumed that the SN magnetic field and the electron fraction do not
change after the shock arrival.
We should note that the magnetic field and the electron fraction after
the shock propagation are still uncertain.
They would change complicatedly.
The distance from the SN was set to be 10 kpc.

We assumed the time evolution of the neutrinos emitted from the neutrino
sphere as follows.
The neutrino flux decreases exponentially
with the time scale of $\tau_\nu = 3$ s.
The total neutrino energy is set to be $3 \times 10^{53}$ ergs,
corresponding to the binding energy of a 1.4 $M_\odot$ neutron star
(e.g., \cite{Yoshida09}).
The neutrino energy is equipartitioned to each flavor of neutrinos.
We considered two spectral types of the neutrino energy spectra, which obey 
Fermi distributions.
The neutrino temperatures in Spectral Type 1 were set to be
$(T_{\nu_e}, T_{\bar{\nu}_e}, T_{\nu_{\mu,\tau}})$ =
(3.2 MeV, 5 MeV, 6 MeV), where $T_{\nu_e}$, $T_{\bar{\nu}_e}$, and
$T_{\nu_{\mu,\tau}}$ mean the temperatures of $\nu_e$, $\bar{\nu}_e$, and
$\nu_{\mu,\tau}$ and $\bar{\nu}_{\mu,\tau}$, respectively 
\cite{Yoshida09,Yoshida06}.
The neutrino chemical potential of each flavor was set to be zero.
In Spectral Type 2, the temperatures and chemical potentials of the neutrinos 
were set as $(T_{\nu_e}, T_{\bar{\nu}_e}, T_{\nu_{\mu,\tau}})$ =
(3.5 MeV, 4 MeV, 7 MeV) and 
$(\mu_{\nu_e}, \mu_{\nu_{\bar{\nu}_e}}, \mu_{\nu_{\mu,\tau}})$ =
(7.4 MeV, 10 MeV, 0 MeV).
This set corresponds to the neutrino spectra of the Livermore model
\cite{Kawagoe10,Totani98}.

In the case of inverted mass hierarchy, neutrino self-interactions
change neutrino-flavors in the innermost region inside RSF-H resonance.
We considered the flavor conversion by neutrino interactions in inverted
mass hierarchy.
The energy splitting in neutrino spectra was taken into account in accordance
with \cite{Fogli07}.
Electron neutrinos above the critical energy $E_C$ are converted from 
$\nu_{\mu,\tau}$.
In the case of anti-neutrinos, $\bar{\nu}_e$ is converted from 
$\bar{\nu}_{\mu,\tau}$ approximately in all energy range.
The critical energy $E_C$ is 8.16 MeV and 7.55 MeV for Spectral
Types 1 and 2, respectively.

We set the neutrino oscillation parameters in accordance with the observational
constraints \cite{SNO04,SK06,Kamland08}.
The squared mass differences $\Delta m_{ij}^2 \equiv m_i^2 - m_j^2$ were 
set to be $\Delta m_{21}^2 = 7.6 \times 10^{-5}$ eV$^2$ and
$|\Delta m_{31}^2| = 2.5 \times 10^{-3}$ eV$^2$.
The mixing angles $\theta_{12}$ and $\theta_{23}$ were set to be
$\sin^2 2\theta_{12} = 0.87$ and $\sin^2 2 \theta_{23} = 1$.
The mixing angle $\theta_{13}$ was set to be $\sin^2 2 \theta_{13} = 0.04$ as
a large value corresponding to an adiabatic MSW-H resonance and
$\sin^2 2 \theta_{13} = 10^{-6}$ as a small value corresponding to 
a non-adiabatic MSW-H resonance.
Mass hierarchy was also considered as a parameter.
When the RSF conversion was taken into account, the strength of the neutrino 
transition magnetic moment was set to be $ 10^{-12} \mu_{\rm B}$.
We numerically solved the neutrino flavor change in accordance with 
\cite{Yoshida09}.

\subsection{Neutrino detection}

We considered the neutrino detection by a liquid Ar detector.
The fidual volume of the liquid Ar detector was assumed to be 100 kton.
This is about 30 times larger than the volume of ICARUS liquid Ar detector.
The detection efficiency of electrons and positrons is assumed to be unity
when the energy of electrons and positrons is larger than $E_{th} = 5$ MeV
and otherwise zero.
The energy resolution was taken as
$\Delta(E_e) = 0.11 \sqrt{E_e ({\rm MeV})} + 0.025 E_e$ MeV 
\cite{Amoruso04}.
In this study, we evaluated the total event numbers of the CC reactions
$^{40}$Ar($\nu_e,e^-)$, $^{40}$Ar($\bar{\nu}_e,e^+)$, and the NC reactions
$^{40}$Ar($\nu,\nu')$.
The rates of the neutrino events and electron and positron energy spectra of 
them were evaluated using the same procedure described in \cite{Yoshida09}.
The cross sections of the neutrino reactions with $^{40}$Ar were taken
from \cite{Gil-Botella03}.

\begin{table}
\caption{
The final neutrino flux of $\nu_e$ [$\phi_{\nu_e}$(fin)] and $\bar{\nu}_e$ 
[$\phi_{\bar{\nu}_e}$(fin)] with the relation of the original neutrino flux 
$\nu_\alpha$ [$\phi_{\nu_\alpha}$].
The neutrino flux $\phi_{\nu_x}$ means the original flux of $\nu_{\mu}$,
$\nu_{\tau}$, $\bar{\nu}_{\mu}$, $\bar{\nu}_{\tau}$.
The first column is model of neutrino oscillation parameters:
the adiabaticity of the RSF-H resonance (R or M), the adiabaticity of the 
MSW-H resonance (A or N), and the mass hierarchy (N or I).
For example, RAN means adiabatic RSF-H resonance, adiabatic MSW-H resonance,
and the normal mass hierarchy.
In inverted mass hierarchy, the $\nu_e$ flux above the critical energy $E_C$
is written.
See text for details.
}
\begin{center}
\begin{tabular}{lcccc}
\hline
Model & $\phi_{\nu_e}$(fin) & Pattern & $\phi_{\bar{\nu}_e}$(fin) & Pattern \\
\hline
RAN & 
$\cos^2 \theta_{12} \phi_{\nu_x} + \sin^2 \theta_{12} \phi_{\bar{\nu}_e}$ &
d &
$\phi_{\nu_x}$ & $\bar{\rm d}$ \\
RAI &
$(\cos^2 \theta_{12} \phi_{\nu_e} + \sin^2 \theta_{12} \phi_{\bar{\nu}_e}
+ \phi_{\nu_x})/2$ & a &
$\{(1+\cos^2 \theta_{12}) \phi_{\nu_x} 
+ \sin^2 \theta_{12} \phi_{\bar{\nu}_e}\}/2$
 & $\bar{\rm c}$ \\
RNN &
$\cos^2 \theta_{12} \phi_{\nu_x} + \sin^2 \theta_{12} \phi_{\nu_e}$ & c &
$\phi_{\nu_x}$ & $\bar{\rm d}$ \\
RNI &
$(\cos^2 \theta_{12} \phi_{\nu_e}+ \sin^2 \theta_{12} \phi_{\bar{\nu}_e} 
+ \phi_{\nu_x})/2$ &
a &
$\{(1+\cos^2 \theta_{12}) \phi_{\nu_x} + \sin^2 \theta_{12} \phi_{\bar{\nu}_e}
\}/2$ & $\bar{\rm c}$ \\
MAN & $\phi_{\nu_x}$ & e &
$\cos^2 \theta_{12} \phi_{\bar{\nu}_e} + \sin^2 \theta_{12} \phi_{\nu_x}$ &
$\bar{\rm a}$ \\
MAI & $\{\cos^2 \theta_{12} \phi_{\nu_e} 
+(1+\sin^2 \theta_{12}) \phi_{\nu_x}\}/2$ & b &
$(\phi_{\bar{\nu}_e} + \phi_{\nu_x})/2$ & $\bar{\rm b}$ \\
MNN &
$\cos^2 \theta_{12} \phi_{\nu_x} + \sin^2 \theta_{12} \phi_{\nu_e}$ & c &
$\cos^2 \theta_{12} \phi_{\bar{\nu}_e} + \sin^2 \theta_{12} \phi_{\nu_x}$ &
$\bar{\rm a}$ \\
MNI & 
$\{\cos^2 \theta_{12} \phi_{\nu_e} + (1+\sin^2 \theta_{12}) \phi_{\nu_x}\}/2$ &
b &
$\{(1+\cos^2 \theta_{12}) \phi_{\nu_x} 
+ \sin^2 \theta_{12} \phi_{\bar{\nu}_e}\}/2$ & $\bar{\rm c}$ \\
\hline
\end{tabular}
\end{center}
\end{table}

\section{Results}

Flavor changes by the RSF conversions and the MSW effects are very 
complicated.
The neutrino flavors are converted in accordance with schematic diagram 
of effective masses of neutrinos and antineutrinos in Fig. 1
(see also \cite{Ando03c,Akhmedov03}).
This figure also indicates the locations of three RSF resonances, i.e., 
RSF-H, RSF-X, and RSF-L (e.g., \cite{Akhmedov03,Yoshida09}) and 
two MSW resonances, i.e., MSW-H and MSW-L (e.g., \cite{ds00}).
The adiabaticities of these resonances depend on the electron-number density
gradient, neutrino energy, the mixing angles, the mass difference, and
the stellar magnetic field (e.g., \cite{Ando03a,Akhmedov03,ds00}).

\begin{table}
\caption{
The event numbers of CC $^{40}$Ar($\nu_e,e^-)$, 
CC $^{40}$Ar($\bar{\nu}_e,e^+)$, and NC $^{40}$Ar($\nu,\nu')$ reactions
with the relation to models of neutrino oscillation parameters.}
\begin{center}
\begin{tabular}{lcccccccc}
\hline
Reaction & RAN & RAI & RNN & RNI & MAN & MAI & MNA & MNI \\
\hline
\multicolumn{9}{c}{Spectral Type 1: 
($T_{\nu_e}$, $T_{\bar{\nu}_e}$, $T_{\nu_{\mu,\tau}}$)
=(3.2 MeV, 5 MeV, 6 MeV), 
($\mu_{\nu_e}$, $\mu_{\bar{\nu}_e}$, $\mu_{\nu_{\mu,\tau}}$)
=(0 MeV, 0 MeV, 0 MeV)} \\
\hline
CC $^{40}$Ar($\nu_e,e^-)$ & 
16384 & 11970 & 12376 & 11981 & 16391 & 12585 & 12312 & 12541 \\
CC $^{40}$Ar($\bar{\nu}_e,e^+)$ &
762 & 763 & 766 & 819 & 580 & 672 & 578 & 823 \\
NC $^{40}$Ar($\nu,\nu')$ &
13668 & 13734 & 13663 & 13733 & 13805 & 13838 & 13805 & 13805 \\
\hline
\multicolumn{9}{c}{Spectral Type 2:
($T_{\nu_e}$, $T_{\bar{\nu}_e}$, $T_{\nu_{\mu,\tau}}$)
=(3.5 MeV, 4 MeV, 7 MeV), 
($\mu_{\nu_e}$, $\mu_{\bar{\nu}_e}$, $\mu_{\nu_{\mu,\tau}}$)
=(7.4 MeV, 10 MeV, 0 MeV)} \\
\hline
CC $^{40}$Ar($\nu_e,e^-)$ & 
19157 & 15534 & 16656 & 15534 & 21908 & 16950 & 16579 & 16981 \\
CC $^{40}$Ar($\bar{\nu}_e,e^+)$ &
1105 & 1119 & 1114 & 1289 & 603 & 866 & 596 & 1294 \\
NC $^{40}$Ar($\nu,\nu')$ &
19063 & 19255 & 19047 & 19253 & 19463 & 19464 & 19464 & 19463 \\
\hline
\end{tabular}
\end{center}
\end{table}

We evaluated the final flux of $\nu_e$ and $\bar{\nu}_e$ as a function
of the initial neutrino flux.
The relation of the final flux of $\nu_e$ and $\bar{\nu}_e$ to
the initial neutrino flux is listed in Table 1.
Here we assumed that $|U_{e1}|^2 = \cos^2\theta_{12}$,
$|U_{e2}|^2 = \sin^2\theta_{12}$, and $|U_{e3}|^2 = 0$, respectively,
for simplicity.
We also assumed that the resonances of RSF-X and RSF-L are non-adiabatic 
and the MSW-L resonance is adiabatic.
We note in inverted mass hierarchy that the neutrino flux above the critical 
energy $E_{\rm C}$ is converted as
$(\phi_{\nu_e}, \phi_{\nu_\mu}, \phi_{\nu_\tau}) \rightarrow
((\phi_{\nu_\mu}+\phi_{\nu_\tau})/2, (\phi_{\nu_e}+\phi_{\nu_\mu})/2,
(\phi_{\nu_e}+\phi_{\nu_\tau})/2)$ and antineutrino flux is converted as
$(\phi_{\bar{\nu}_e}, \phi_{\bar{\nu}_\mu}, \phi_{\bar{\nu}_\tau}) \rightarrow
((\phi_{\bar{\nu}_\mu}+\phi_{\bar{\nu}_\tau})/2, 
(\phi_{\bar{\nu}_e}+\phi_{\bar{\nu}_\mu})/2,
(\phi_{\bar{\nu}_e}+\phi_{\bar{\nu}_\tau})/2)$ by neutrino self-interactions
before the RSF conversions.

In Table 1, there are eight models of neutrino oscillation parameters:
the adiabaticities of the RSF-H and MSW-H resonances and the mass hierarchies.
We denote these models as three characters.
The first character means the adiabaticity of the RSF-H resonance;
^^ ^^ R" means that the RSF-H is adiabatic, i.e., the RSF conversion is 
effective, and ^^ ^^ M" means that the RSF-H resonance is non-adiabatic
thus the RSF conversion does not work.
The second character means the adiabaticity of the MSW-H resonance; 
^^ ^^ A" means adiabatic and ^^ ^^ N" means non-adiabatic.
The third character means the mass hierarchy; ^^ ^^ N" and ^^ ^^ I" correspond
to ^^ ^^ normal" and ^^ ^^ inverted" mass hierarchies, respectively.

Generally, the temperatures of SN neutrinos indicate the relation of
$T_{\nu_e} < T_{\bar{\nu}_e} < T_{\nu_{\mu,\tau}}$.
In this case the event number of $\nu_e$, $N_{\nu_e}$, or 
$\bar{\nu}_e$, $N_{\bar{\nu}_e}$, becomes large when
$\nu_e$ or $\bar{\nu}_e$ is converted from energetic $\nu_{\mu,\tau}$.
From Table 1, we expect the event number of $\nu_e$ increases with 
[pattern a] $N_{\nu_e}$(RAI), $N_{\nu_e}$(RNI) $<$ 
[pattern b] $N_{\nu_e}$(MAI), $N_{\nu_e}$(MNI) $<$
[pattern c] $N_{\nu_e}$(RNN), $N_{\nu_e}$(MNN) $<$
[pattern d] $N_{\nu_e}$(RAN) $<$ [pattern e] $N_{\nu_e}$(MAN) 
and that the event number of $\bar{\nu}_e$ increases with 
[pattern $\bar{\rm a}$] $N_{\bar{\nu}_e}$(MAN), $N_{\bar{\nu}_e}$(MNN) $<$ 
[pattern $\bar{\rm b}$] $N_{\bar{\nu}_e}$(MAI) $<$ 
[pattern $\bar{\rm c}$] $N_{\bar{\nu}_e}$(RAI), $N_{\bar{\nu}_e}$(RNI), 
$N_{\bar{\nu}_e}$(MNI) $<$ 
[pattern $\bar{\rm d}$] $N_{\bar{\nu}_e}$(RAN), $N_{\bar{\nu}_e}$(RNN).

Considering the above relations we numerically evaluated the total event 
numbers of the CC reactions $^{40}$Ar($\nu_e,e^-)$ and 
$^{40}$Ar($\bar{\nu}_e,e^+)$ and the NC reactions 
$^{40}$Ar($\nu,\nu')$ with the two SN neutrino spectral types.
Table 2 shows the total event numbers of these three reactions.
The event numbers of $^{40}$Ar($\nu_e,e^-)$ and the NC reactions are of
order $10^4$.
The event number of $^{40}$Ar($\bar{\nu}_e,e^+)$ is in the range of hundreds 
to 1300.
The event numbers of the CC reactions depend on oscillation parameters and 
neutrino magnetic moment.
However, the NC event scarcely depends on.
When a model of oscillation parameters is given, the event numbers 
in Spectral Type 2 are larger than the corresponding numbers in
Spectral Type 1.
This relation is due to the fact that the average energies of $\nu_e$, 
$\bar{\nu}_e$, and $\nu_{\mu,\tau}$ in Spectral Type 2 are larger than the
corresponding average energies in Spectral Type 1.

Next, we indicate the dependence of the event number ratio of
$^{40}$Ar($\nu_e,e^-)$ reaction to the NC reactions 
($r_{{\rm CC}(\nu_e)/{\rm NC}}=N_{\nu_e}/N_{{\rm NC}}$) and that of
$^{40}$Ar($\bar{\nu}_e,e^+)$ reaction to the NC reactions 
($r_{{\rm CC}(\bar{\nu}_e)/{\rm NC}}=N_{\bar{\nu}_e}/N_{{\rm NC}}$) 
on the oscillation parameters.
Figure 2(a) shows the relation of the two event number ratios in 
Spectral Type 1.
The event number ratios on this figure do not overlap with two exceptions.
The ratios of models RNN and MNI overlap with error bars.
The ratios of models RAI and RNI also do.

\begin{figure}
\begin{center}
\includegraphics[angle=-90,width=7.5cm]{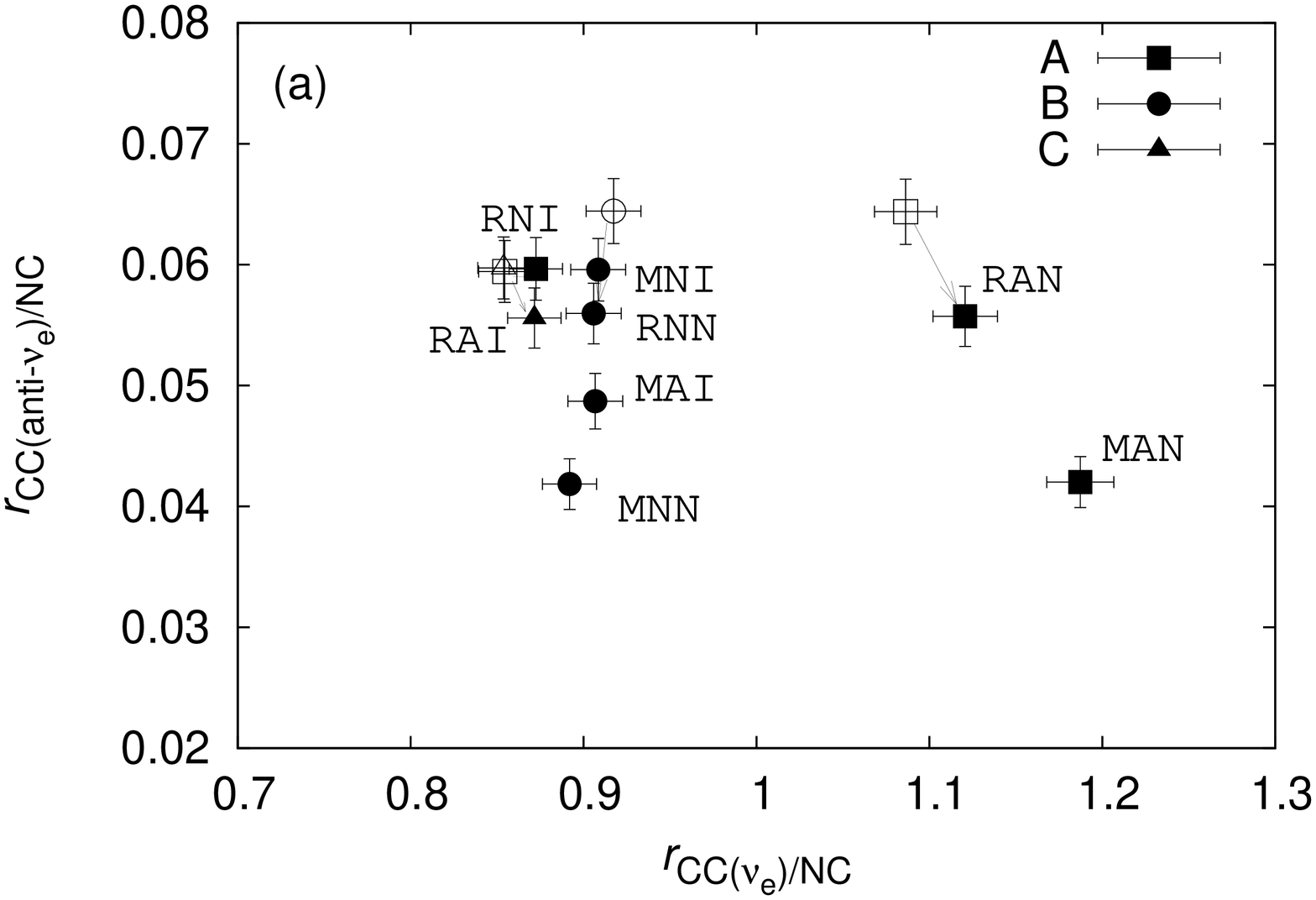}
\includegraphics[angle=-90,width=7.5cm]{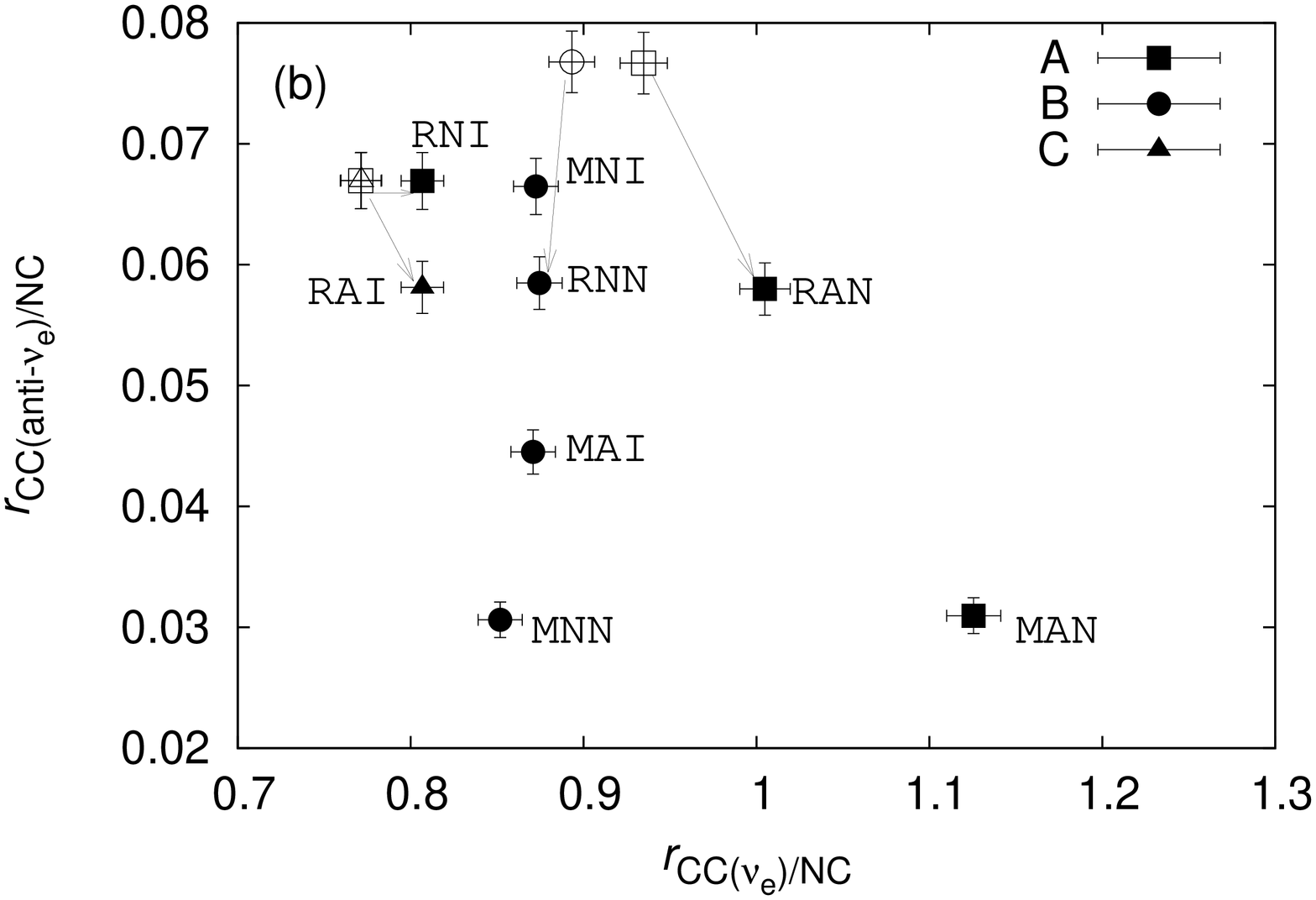}
\end{center}
\caption{
The relation between the event number ratio of 
$^{40}$Ar($\nu_e,e^-)$ to the NC reactions of $^{40}$Ar 
to the NC reactions of $^{40}$Ar $r_{{\rm CC}(\nu_e)/{\rm NC}}$ 
and the ratio of $^{40}$Ar($\bar{\nu}_e,e^+)$ to the NC reactions of $^{40}$Ar
$r_{{\rm CC}(\bar{\nu}_e)/{\rm NC}}$.
Panels (a) and (b) are the number ratios in Spectral Types 1 and 2,
respectively.
Squares, circles, and triangles indicate patterns A, B, and C, respectively, 
of the observation of neutronization burst (see text for more details).
Open symbols indicate the corresponding ratios when the change of the 
adiabaticities in the RSF-H and MSW-H resonances by the shock propagation 
in the SN is not taken into account.
}
\end{figure}

The event number ratios in Fig. 2(a) do not obey the inequalities indicated 
above.
This is due to the shock effect on the RSF conversions.
When the shock wave passes through an adiabatic resonance, the adiabaticities
changes to non-adiabatic.
Since the RSF-H resonance is located in deep region of the SN ejecta, the
adiabaticity of the resonance changes when neutrino irradiation is
still strong.
As a result, the shock effect at the RSF-H resonance is stronger than
that at the MSW-H resonance in the total neutrino event.

We also show the event ratios as open symbols in models RAN, RAI, RNA,
and RNI without adiabaticity change by the shock passage of the RSF-H
resonance.
The ratios with closed symbols in the four models are different from the 
corresponding ratios with relations of the event numbers estimated in Table 1.
When the shock effect is not taken into account, the numerical
result is consistent with the theoretical estimation.
For example, the event ratio for $\nu_e$ increases with 
[pattern a] $N_{\nu_e}$(RAI), $N_{\nu_e}$(RNI) $<$ 
[pattern b] $N_{\nu_e}$(MAI), $N_{\nu_e}$(MNI) and 
[pattern c] $N_{\nu_e}$(RNN), $N_{\nu_e}$(MNN) $<$ 
[pattern d] $N_{\nu_e}$(RAN) $<$ [pattern e] $N_{\nu_e}$(MAN).

When the RSF conversions are effective, the event ratios with shock effect 
are different from those without shock effect.
On the other hand, the ratios with ineffective RSF conversion cases
hardly depend on the shock effect.
This is because the time when the shock arrives at the MSW-H resonance
is later than the shock arrival time to the RSF-H resonance and most
of neutrinos have passed until the shock arrival at the MSW-H resonance.
If we obtain large difference between the event ratios for $\nu_e$ and 
$\bar{\nu}_e$ with the shock effect and those without the shock effect, 
we can discuss the possibility of the RSF conversions.

Neutronization burst is another special neutrino signal appearing 
at the core bounce.
The spectra of the neutronization burst are quite different from 
the total neutrino signal and the main component of the neutronization
burst is $\nu_e$.
The neutrinos in the neutronization burst are scarcely affected
by neutrino self-interactions.
Most of the released neutrinos are $\nu_e$ in neutronization burst, 
so that the interaction of $\nu_e$ and other flavor of neutrinos and 
antineutrinos is quite small.
The neutrino signal depends on neutrino oscillation parameters.
The effects of RSF conversions and the MSW effects on $\bar{\nu}_e$ have been 
investigated in \cite{Ando03a,Ando03c}.
We investigate both $\nu_e$ and $\bar{\nu}_e$ signals 
in the neutronization burst.
Figure 2 shows three patterns of the detection of 
the neutronization burst.
Table 3 indicates the final flux of $\nu_e$ and $\bar{\nu}_e$.
Pattern A corresponds to no signals of neutronization burst.
Pattern B corresponds to the $\nu_e$ detection and Pattern C indicates
$\bar{\nu}_e$ detection in neutronization burst.
The signal of the neutronization burst distinguishes the neutrino 
events between model RAI and model RNI.
Thus, the $\nu_e$ and $\bar{\nu}_e$ signals from the neutronization
burst also constrain the neutrino oscillation parameters and neutrino 
magnetic moment.

Figure 2(b) shows the relations of the two event number ratios in Spectral
Type 2.
We see that the relation of the event number ratios is similar to Fig. 2(a);
the dependence of the event number ratios on oscillation parameters
is larger.
This is due to larger difference between the $\nu_{\mu,\tau}$ temperature
and those of $\nu_e$ and $\bar{\nu}_e$.

\begin{table}
\caption{
The final neutrino flux of $\nu_e$ [$\phi_{\nu_e}$(fin)] and $\bar{\nu}_e$ 
[$\phi_{\bar{\nu}_e}$(fin)] with the relation of the original neutrino flux 
$\nu_\alpha$ [$\phi_{\nu_\alpha}$] in the neutronization burst.
The first column is model of neutrino oscillation parameters as described
in Table 1.
}
\begin{center}
\begin{tabular}{lccc}
\hline
Model & $\phi_{\nu_e}$(fin) & $\phi_{\bar{\nu}_e}$(fin) & Pattern \\
\hline
RAN & 0 & 0 & A \\
RAI & 0 & $\cos^2 \theta_{12} \phi_{\nu_e}$ & C \\
RNN & $\sin^2 \theta_{12} \phi_{\nu_e}$ & 0 & B \\
RNI & 0 & 0 & A \\
MAN & 0 & 0 & A \\
MAI & $\sin^2 \theta_{12} \phi_{\nu_e}$ & 0 & B \\
MNN & $\sin^2 \theta_{12} \phi_{\nu_e}$ & 0 & B \\
MNI & $\sin^2 \theta_{12} \phi_{\nu_e}$ & 0 & B \\
\hline
\end{tabular}
\end{center}
\end{table}

If we can expect the initial neutrino spectra precisely, we would constrain
the neutrino oscillation parameters and neutrino magnetic moment from
SN neutrino signals detected by a liquid Ar detector.
The total events of the CC reactions of $\nu_e$ and $\bar{\nu}_e$ and the
NC reactions will indicate different signals by different oscillation
parameters for most cases.
The shock effect in the total CC events of $\nu_e$ and $\bar{\nu}_e$ will
clarify whether the RSF conversions are effective or not.
This suggests the upper limit of the product of neutrino magnetic moment
and the SN magnetic field.
The detection of neutronization burst signal by $\nu_e$ or 
$\bar{\nu}_e$ detector or no detection gives additional constraint to 
the oscillation parameters.

Even if the initial neutrino spectra are expected less precisely, we would
constrain some oscillation parameters and neutrino magnetic moment.
Large and small $\bar{\nu}_e$ events with large $\nu_e$ events and no
detection of neutronization burst should be the evidence for models 
RAN and MAN, respectively.
Small $\nu_e$ event, large $\bar{\nu}_e$ event, and no detection of
neutronization burst will indicate the evidence for model RNI.
As first discussed in \cite{Ando03c}, the detection of 
neutronization burst in $\bar{\nu}_e$ signal is 
a strong evidence for model RAI.
Small $\nu_e$ event will confirm this evidence.
Small $\bar{\nu}_e$ event with the detection of the neutronization
burst in $\nu_e$ signal will exclude the possibility of large neutrino 
magnetic moment enough to bring about the RSF conversions.

We have shown that the shock effect appears in the event number ratios of
$\nu_e$ and $\bar{\nu}_e$ signals when the RSF conversions are effective.
However, if a SN explodes jet-like, different time dependence will be
seen by the adiabaticity change of the MSW-H resonance \cite{Kawagoe09}.
Magneto-hydrodynamic simulations of SNe indicated that magnetically-driven
jet explosion is induced by strong toroidal magnetic field 
\cite{tk09}.
Under such explosions the points and adiabaticities of the RSF and MSW
resonances would be quite different from those in spherical explosions.
Thus, the investigation of the neutrino signal from magnetically-driven
SNe taking into account the RSF conversion is fascinated.
Analyses with optical observations of SNe \cite{mk08} help distinguishing 
the effects by the RSF conversions and jet-like explosion of the SN.

\section{Conclusions}

We have investigated traces of large neutrino magnetic moment, the RSF
conversions, in SN neutrino signals observed by a 100 kton liquid Ar detector.
We predict that different signals will be observed in the event number ratios
of the CC $\nu_e$ and $\bar{\nu}_e$ reactions to the NC reactions with
different oscillation parameters in both normal and inverted mass hierarchies.
Large shock effect will be found in the total event number ratios
when the RSF conversions are effective.
Combined with the signals of the neutronization burst, the $\nu_e$ 
and $\bar{\nu}_e$
signals would constrain neutrino oscillation parameters and the neutrino
magnetic moment.
The observations of both $\nu_e$ and $\bar{\nu}_e$ will open the
possibility for constraining neutrino oscillation parameters as well as
neutrino magnetic moment.

T.Y. thanks Mary Bishai and Kate Scholberg for kindly providing useful
information of liquid Ar detectors in LBNE Project.
Numerical computations were in part carried out on the general-purpose
PC farm at Center for Computational Astrophysics, CfCA, of
National Astronomical Observatory of Japan.
This work has been supported in part by the Ministry of Education,
Culture, Sports, Science and Technology, Scientific Research 
(C)(20540284, 23540287).



%% References


\bibliographystyle{model1-num-names}
%\bibliography{rsf2}



\end{document}